# Statistical Timing Based Optimization using Gate Sizing

Aseem Agarwal

University of Michigan,
Ann Arbor, MI
abagarwa@umich.edu

Kaviraj Chopra

University of Michigan,
Ann Arbor, MI
kaviraj@umich.edu

David Blaauw

University of Michigan,
Ann Arbor, MI
blaauw@umich.edu

**Abstract**

*The increased dominance of intra-die process variations has motivated the field of Statistical Static Timing Analysis (SSTA) and has raised the need for SSTA-based circuit optimization. In this paper, we propose a new sensitivity based, statistical gate sizing method. Since brute-force computation of the change in circuit delay distribution to gate size change is computationally expensive, we propose an efficient and exact pruning algorithm. The pruning algorithm is based on a novel theory of perturbation bounds which are shown to decrease as they propagate through the circuit. This allows pruning of gate sensitivities without complete propagation of their perturbations. We apply our proposed optimization algorithm to ISCAS benchmark circuits and demonstrate the accuracy and efficiency of the proposed method. Our results show an improvement of up to 10.5% in the 99-percentile circuit delay for the same circuit area, using the proposed statistical optimizer and a run time improvement of up to 56x compared to the brute-force approach.*

## 1 Introduction

Static Timing Analysis (STA) has been the mainstay of performance verification for the past two decades. Traditionally, process variation has been addressed in STA using corner-based analysis where all gates are assumed to operate at a worst-, typical- or best-case condition and within-die variability is not modeled. However, in the nanometer regime, within-die variation has become a substantial portion of the overall variability and corner-based STA suffers from significant inaccuracy. This has given rise to a new field of statistical timing analysis known as SSTA.

In SSTA, the circuit delay is considered a random variable and the objective of SSTA is to compute its probability distribution. So-called *block-based* SSTA approaches [1-5] are analogous to STA in that they propagate arrival times through the circuit. As the arrival times traverse gates, the delay of the gate is added to the arrival time and a maximum arrival time is selected when multiple arrival times converge at a gate. In SSTA, the arrival times also become random variables, and hence, the addition and maximum operations of STA are replaced by convolution and a statistical maximum, respectively. Like STA, they require a single pass of the circuit to compute the circuit delay distribution. From the CDF (cumulative distribution function) of the circuit delay, the user is then able to obtain the percentage of fabricated dies which meets a certain delay requirement, or conversely, the expected performance for a particular yield. In turn, gate or transistor sizing approaches should consider such metrics for their objective function and should perform their optimization in a statistically aware manner.

SSTA-based optimization can significantly improve the yield of a design compared to deterministic optimization. This is due to the fact that deterministic optimization tends to create a so-called "wall" of critical and nearly critical paths, as shown in Figure 1a, since there is no incentive to improve path delays that are not critical. All critical paths can affect the circuit delay due to delay variability, and hence, a balanced circuit with many near-critical paths is highly susceptible to process variations. This is illustrated in Figure 1(a) and (b), where a balanced and unbalanced path distributions are shown with their associated circuit delay distributions. While both path distributions have the same deterministic circuit delay, the unbalanced distribution results in a better statistical circuit delay since it has fewer near-critical paths. Hence, deterministic optimization can actually worsen the true statistical circuit delay due to the lack of a true statistical objective function.

Recently, a number of statistical optimization algorithms have been proposed in [6-9]. In [6] it is shown that by performing deterministic optimization, it is possible to degrade the performance of the die statistically due to the creation of a timing wall (Figure 1). Hence, in [7] the authors propose a method to avoid the formation of such a wall by purposely improving non-critical paths in the deterministic optimization. In [8] and [9], the statistical optimization problem has been considered as a nonlinear programming problem. Delays are considered to be gaussian and approximations are used for computing the statistical maximum. In [9], a heuristic approach is proposed using the concept of statistically 'undominated' paths. These approaches suffer from lack of true sensitivity computation and prohibitive runtimes for large circuits.

In this paper, we therefore propose a new sensitivity based, statistical gate sizing algorithm. We use a coordinate descent algorithm where in each iteration the gate with the highest sensitivity is sized up. We show that such a statistically aware optimization can improve the 99-percentile delay by up to 10.5% over that obtained with traditional deterministic optimization. It should be noted, as shown in Figure 2, that a sizing change can impact both the mean and the shape of the circuit delay CDF. Depending on the objective specified by the user, the CDF perturbation can be evaluated in a number of ways. In this paper, we consider as objective the CDF delay at the 99% probability or confidence point, as shown in Figure 2. Hence, the computed sensitivity is measured as the change in the 99-percentile delay of the circuit delay CDF. However, other objective functions could be equally well supported by the proposed framework.

Since brute-force computation of such CDF perturbation sensitivities is extremely expensive, a key contribution of this paper is an efficient and exact pruning algorithm that allows for identification of the most sensitive gate in the circuit. Our pruning approach is based on a proposed theory of bounds on CDF perturbations due to sizing. We establish the useful property that these perturbation bounds can only diminish as the arrival time perturbations are propagated through the circuit using convolutions and maximum operations. Based on this property, we propose a pruning algorithm for finding the highest sensitivity gate in a sizing iteration, without complete propagation of the perturbed arrival times for all gates. We perform an iterative propagation of perturbed arrival times for a pruned set of gates by maintaining their so-called perturbation fronts (defined later). We test our approach on a number of benchmark circuits, and demonstrate up to 56 times faster runtimes than the brute force sensitivity computation based optimization, without loss in accuracy.

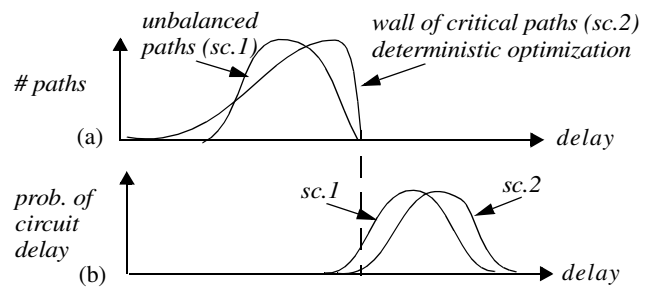

**Figure 1. a) distribution of paths in a circuit b) corresponding circuit delay PDFs**

The remainder of this paper is organized as follows. In Section 2, we present our modeling assumptions, along with the problem formulation, basic definitions and delay model. In Section 3, we present our approach for sensitivity computation and optimization. In Section 4, we present our results and compare deterministic optimization with brute force statistical and our proposed accelerated approach. Finally, in Section 5 we draw our conclusions.

## 2 Problem Formulation

In this section we define our modeling assumption and our SSTA approach. We also formulate the statistical optimization problem and present basic definitions and the delay model.

Gate delay variability is composed of two primary components: inter-die (between-die) variability and intra-die (within-die) variability. Inter-die variability expresses the change in gate delay from one die to the next and has traditionally been modeled using corner analysis with reasonable accuracy. The main focus of SSTA has therefore been on intra-die variability, which corner-based analysis is unable to model. Hence, similar to the optimization approaches proposed in [8,9] we focus on intra-die variability in this paper.

One of the difficulties in SSTA arises from reconvergent circuit structures, which results in correlations between arrival times. In [2,3], it was shown that the worst-case runtime for exact computation of the circuit delay CDF in a reconvergent circuit is exponential with circuit size, making its computation impractical. However, in [3] a simple method where these correlations are ignored was shown to result in an upper bound on the circuit delay CDF and hence a conservative analysis. Furthermore, it was shown that these bounds are typically tight and give a reasonably close approximation of the exact circuit delay CDF while their computation runtime is linear with circuit size. In this paper, we therefore use the bounds proposed in [3] for computation of the circuit delay CDF. It is important to note that the optimization objective is defined on this bound of the circuit delay CDF and not on the exact circuit delay CDF itself, since this would lead to prohibitive runtimes. However, we show in the result section that the optimization of the bounds, as performed by our method, results in nearly equivalent improvement of the exact circuit delay, as verified using Monte-Carlo simulation.

In addition to reconvergent circuit structures, spatial correlation of the gate delays can also give rise to correlation of arrival times [5]. However, similar to previous optimization methods [8,9], our optimization approach does not model such correlations at this time, although the proposed methods form a basis from which such correlations can be incorporated.

In a statistical timing paradigm, the delay of the circuit is a random variable. As a result, one needs to determine an appropriate objective function for optimization, defined on the distribution of the circuit delay random variable. Since we use propagation of discretized arrival time PDFs, and not merely the statistical measures such as mean and variance, we obtain the entire shape of the circuit delay distribution. Hence, the proposed framework can support a wide range of cost functions as optimization objectives. For simplicity of explanation, however, we choose as our optimization objective the $p$-percentile point $T(p)$ of the delay distribution. In our experiments, we choose $p$ to be the 99-percentile point, as shown in Figure 2.

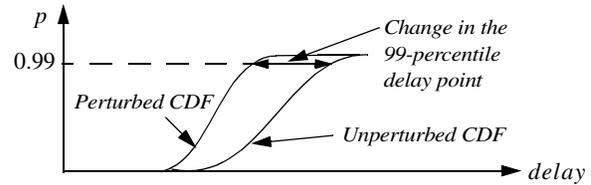

**Figure 2. Optimization objective (99-percentile delay point)**

We use the following graph representation for our circuits.

**Definition 1.** A *timing graph* $G$ is a directed graph having exactly one source and one sink node: $G=\{N,E,ns,nf\}$, where $N=\{n_1,n_2,...,n_k\}$ is a set of nodes, $E=\{e_1,e_2,...,e_l\}$ is a set of edges, $ns \in N$ is the source node, and $nf \in N$ is the sink node and each edge $e \in E$ is simply an ordered pair of nodes $e=(n_i,n_j)$.

The nodes in the timing graph correspond to nets in the circuit, and the edges in the graph correspond to connections from gate inputs to gate outputs.

### 2.1 Delay Model

We use a simple delay model for our experiments, similar to that used in [6], based on the logic effort model. In this model, the pin-to-pin delay (edge delay) of a gate is defined as

$$D_e = D_{int} + K \times C_{load}/C_{cell}, \qquad \text{(EQ 1)}$$

where, $D_{int}$ is a constant, intrinsic delay due to cell-internal capacitances, $C_{load}$ is the total load capacitance, $K$ is a constant for the standard cell and $C_{cell}$ is the total capacitance of the standard cell.

We determine these constants for all the standard cells in our synthesis library for our experiments. For the statistical modeling of these delays we assume that the standard deviation is a fixed percentage of the nominal delay, although our method is not restricted to this model.

## 3 Proposed Optimization Approach

We first present in Section 3.1 the straightforward approach to performing statistical optimization using sensitivities. In Section 3.2, we develop novel properties of sensitivity propagation based on which an efficient pruning algorithm is presented in Section 3.3.

### 3.1 Straightforward Approach

Our brute force statistical algorithm is similar in structure to a deterministic coordinate descent algorithm. The deterministic optimization is sensitivity based and iteratively minimizes the circuit delay starting from a minimum size implementation. During each iteration, the gate with the maximum sensitivity is identified and sized up. If the optimization is deterministic in nature, any gate that improves the circuit delay by being sized up, must lie on the critical path of the circuit and hence, the sensitivity computation can be restricted to only those gates on the critical path.

However, in optimizing statistical circuit delay, there may be no single longest path, because the circuit delay PDF is a combination of all the path delay PDFs. Hence, statistical sensitivity needs to be computed for all gates in the circuit making a sensitivity based statistical optimization significantly more computationally demanding. According to the objective function defined in Section 2, the statistical sensitivity is the change in the *p*-percentile point of the circuit delay CDF due to the upsizing of a gate. This means that the perturbation of sizing a gate must be propagated to the sink node in order to calculate the sensitivity of the gate. This necessitates a statistical timing analysis run for each gate in the circuit at every sizing step of the algorithm with a runtime complexity of O(N*E) for every sizing iteration, where N is the number of nodes and E is the number of edges of graph *G*. This results in unacceptable runtimes. Therefore, we propose an approach where the gate with maximum sensitivity can be identified without explicit propagation of perturbed arrival time CDFs for each gate.

### 3.2 Properties of sensitivity propagation

To allow for pruning of sensitivities, we now introduce the following useful definitions and properties of sensitivity propagation.

As shown in Figure 5, $A_i$ is the CDF of the arrival time random variable at node $i$ and $A'_i$ is the corresponding perturbed CDF obtained by scaling up a gate. Their PDFs are denoted by $a_i$ and $a'_i$, respectively. We define the difference in the *p*-percentile point of the CDFs $A_i$ and $A'_i$ as $\delta_i(p) = T(A_i, p) - T(A'_i, p)$. The maximum difference over all $p$ is given by $\Delta_i = max_p \delta_i(p)$.

First, we assume that the perturbed CDF $A'_i$ has the exact same shape as the unperturbed CDF $A_i$ and differs from $A_i$ only by a constant shift in time, i.e. $A_i(t) = A'_i(t - \Delta_i)$ and also $a_i(t) = a'_i(t - \Delta_i)$. This is assumed to be true for all perturbed CDFs. Under this assumption, we prove in Theorems 1 through 3 that the maximum difference $\Delta_i$ between the perturbed and unperturbed CDFs at a node can not increase as the perturbed CDFs are propagated through the circuit using convolution and statistical maximum. This property is useful in bounding the difference between the perturbed and unperturbed CDFs at the sink node, without complete propagation of the gate's perturbed CDF to the sink node. However, a change in a gate size often affects not only the mean of the gate delay, but also the shape of the CDF. Therefore, we show that it is possible to construct a lower bound on the perturbed CDF, such that the shape of this lower bound is identical to the unperturbed CDF, as illustrated by CDF $B'_i$ in Figure 5. We then apply Theorems 1 through 3 to this lower bound on the perturbed CDF and show that these theorems are true for any shape perturbation. Finally, we show the use of these theorems to effectively prune out the propagation of perturbed CDFs.

**Theorem 1. Convolution operation:** Consider the timing graph shown in Figure 3a. Let $a_i$ and $a'_i$ be the original and the perturbed PDF at node $i$ such that $a_i(t) = a'_i(t - \Delta_i)$ and let $d_e$ be the delay PDF of edge $e$. If the arrival time PDF $a_j$ and the perturbed $a'_j$ at node $j$ are given by $a_j = Conv(a_i, d_e)$ and $a'_j = Conv(a'_i, d_e)$, then $\Delta_i = \Delta_j$.

**Proof** : The proof is obvious and omitted for brevity.

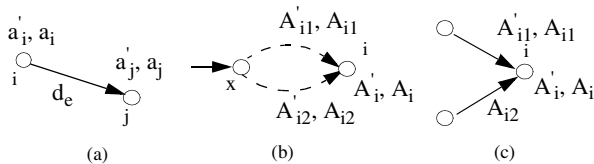

**Figure 3. Timing graphs**

In the following two theorems, we show that a similar property to that of Theorem 1 holds for the maximum operation. As previously mentioned, we assume that correlation of the arrival times due to reconvergent fanout can be ignored for the maximum operation and hence, the theorems are defined for an upper bound of the exact arrival time CDF [3].

**Theorem 2. Max operation with multiple perturbed arrival times:** Consider a node $i$ in the probabilistic timing graph shown in Figure 3b. Let $A_{i1}$ and $A_{i2}$ be the arrival time CDFs of two fanin subgraphs incident at node $i$. Let $A'_{i1}$ and $A'_{i2}$ be the perturbed CDFs obtained by scaling a single gate $x$ that is common to the fanin cones of $A_{i1}$ and $A_{i2}$. If the arrival time CDF $A_i$ and perturbed CDF $A'_i$ at node $i$ are given by, $A_i = max(A_{i1}, A_{i2})$ and $A'_i = max(A'_{i1}, A'_{i2})$ respectively, then $\Delta_i \leq max(\Delta_{i1}, \Delta_{i2})$.

**Proof** : We consider two cases.

**case 1**: Consider $\Delta_{i1} = \Delta_{i2}$.

By definition of maximum operation assuming independence,

$$A_i(t) = A_{i1}(t) \cdot A_{i2}(t) \quad (EQ\ 2)$$

and $A'_i(t - \Delta_{i1}) = A'_{i1}(t - \Delta_{i1}) \cdot A'_{i2}(t - \Delta_{i1}) \quad (EQ\ 3)$

R.H.S. of EQ2 and EQ3 being same, $A_i(t) = A'_i(t - \Delta_{i1})$, but we know that $A_i(t) = A'_i(t - \Delta_i)$. Hence, $\Delta_i = \Delta_{i1} = \Delta_{i2}$.

**case 2**: Consider $\Delta_{i1} \neq \Delta_{i2}$.

Without loss of generality, assume $\Delta_{i1} > \Delta_{i2}$, and also $\Delta_{i1} = \Delta_{i2(case1)}$ and $\Delta_{i2} < \Delta_{i2(case1)}$ as shown in Figure 4. We define a new CDF due to $\Delta_{i2}$ as $A''_{i2}$, and the new resultant max as $A''_i$. Again by definition,

$$A''_i(t - \Delta_{i1}) = A'_{i1}(t - \Delta_{i1}) \cdot A''_{i2}(t - \Delta_{i1}) \quad (EQ\ 4)$$

Also, $A''_{i2}(t - \Delta_{i1}) < A'_{i2}(t - \Delta_{i1})$, because $\Delta_{i2} < \Delta_{i2(case1)}$. Hence, by equating EQ3 and EQ4, we get $A''_i(t - \Delta_{i1}) < A'_i(t - \Delta_{i1})$. This implies, $T(A''_i, p) > T(A'_i, p)$, and $T(A_i, p) - T(A''_i, p) < T(A_i, p) - T(A'_i, p)$ by algebraic manipulation. Hence, $\Delta_i < \Delta_{i1}$. □

Note that the proof can be trivially extended for gates with more than two inputs.

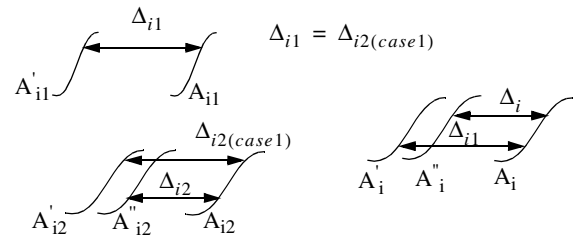

**Figure 4. Arrival time CDFs - max operation (case 2)**

**Theorem 3. Max operation with single perturbed arrival time:** Consider a node $i$ in the timing graph shown in Figure 3c. Let $A_{i1}$ and $A_{i2}$ be the arrival time CDFs of two fanin subgraphs incident at node $i$. Let $A'_{i1}$ be the only perturbed CDF. If the arrival time CDF $A_i$ and perturbed CDF $A'_i$ at node $i$ are given by, $A_i = max(A_{i1}, A_{i2})$ and $A'_i = max(A'_{i1}, A_{i2})$ respectively, then $\Delta_i \leq \Delta_{i1}$.

**Proof** : This is a special case of Theorem 2, where $\Delta_{i2} = 0$.

The above three theorems were defined assuming that the perturbed CDF has the exact same shape as the unperturbed CDF. As mentioned, this may not be true in practice and hence, we define a lower bound on the perturbed CDF which has the exact same shape as the unperturbed CDF as follows.

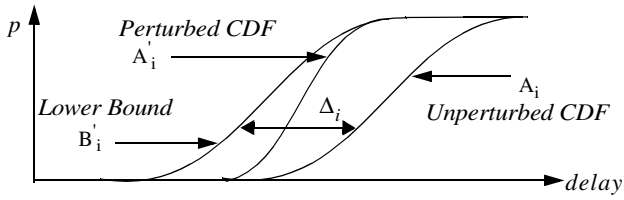

Figure 5. Arrival time CDFs at node i

**Definition 2.** The lower bound CDF $B'_i$ of perturbed arrival time $A'_i$ is defined as the time shifted CDF $A_i$ by $\Delta_i$ (Figure 5).

Since the shape of the lower bound $B'_i$ is the same as that of the unperturbed CDF $A_i$, Theorems 1 through 3 can be applied to this lower bound. Note, however, that the maximum time difference between the *lower bound* of the perturbed CDF $B'_i$ and the unperturbed CDF $A_i$ is equal to the maximum difference between the perturbed CDF $A'_i$ itself, and $A_i$ (by Definition 2). Hence, implicitly, Theorems 1 through 3 also hold for arbitrary shaped perturbations of an arrival time CDF. This allows the use of the perturbation bound $\Delta_i$ as an upper bound on the actual difference between the perturbed and unperturbed CDFs at the sink node. Using this bound allows gates to be pruned from consideration for the highest sensitivity gate as explained in more detail in Section 3.3.

Before presenting such a general upper bound on the perturbation of a CDF at the sink node, we first recognize that when we propagate a perturbed CDF in a circuit, multiple perturbed CDFs are generated at points of multiple fanout. We therefore introduce a so-called perturbation front, $P_k$, which is the set of nodes that is visited in each iteration of a breadth-first propagation of the perturbed CDF to the sink node. We now define the maximum over all $\Delta_i$ where node $i$ belongs to a perturbation front due to upsizing gate $x$ as $\Delta m_x = max_i \Delta_i$. Note that when the perturbation front reaches the sink node, it consists of only a single perturbed arrival time corresponding to the sink node.

**Theorem 4.** Given a perturbation front $P_k$ associated with a gate $x$ then $\Delta_{nf} \leq \Delta m_x$, where $\Delta m_x$ is the maximum difference between perturbed and unperturbed arrival times over all nodes in the perturbation front $P_k$ and $\Delta_{nf}$ is the maximum difference between perturbed and unperturbed CDF at the sink node $nf$.

**Proof** : The proof follows from Theorems 1 through 3.

Theorem 4 states that the maximum difference between the perturbed and unperturbed CDF at the sink node, is bounded by the maximum change of the perturbed and unperturbed CDFs in the perturbation front.

### 3.3 Our Algorithm

In this section, our statistical gate sizing algorithm is presented. The objective function is the $p$-percentile point of the circuit delay CDF. The sensitivity $S_x$ of gate $x$ is computed numerically using the ratio of the change in $p$-percentile circuit delay per unit change in gate width: $S_x = \delta_{nf}(p)/\Delta w_x$ where, $\Delta w_x$ is the change in gate width and $nf$ is the sink node of $G$. Based on the computed sensitivities, the most sensitive gate in each iteration of the coordinate descent is selected.

The goal of the inner loop of the optimization is to find the gate with maximum sensitivity without performing a complete SSTA run for each gate perturbation in the circuit. The idea is to propagate highly sensitive gates (i.e. gates which have a large value of $S_i$) to the sink node and then use their $S_i$ value to prune out gates which can be shown to have a lesser sensitivity using the proposed bounds. Given a gate $x$ with partially propagated arrival time CDFs at perturbation front $P_k$, we define the perturbation front sensitivity bound $Sm_x = \Delta m_x / \Delta w_x$, where $\Delta m_x$ is the maximum perturbation change across the nodes of the front. From Theorem 4 it follows that $Sm_x < S_x$ and hence the sensitivity bound $Sm_x$ can be used to prune gate $x$ before its perturbation front reaches the sink node. In other words, if at any time during the propagation of the front for gate $x$ the bound $Sm_x$ become less than a previously computed sensitivity $S_i$ of gate $i$, gate $x$ can be eliminated from further consideration.

It is advantageous to identify a gate with a high sensitivity value $S_i$ early in the analysis so that a large number of gates can be pruned. In our approach, we therefore perform level by level propagation of perturbed arrival times in an iterative manner. During every iteration the perturbation front with the maximum $Sm_x$ value is propagated one level forward and its $Sm_x$ value is recomputed. When a perturbation front reaches the sink node, its true sensitivity $S_i$ is computed and is used to prune the perturbation front of other gates. The pseudo-code of the statistical gate sizing algorithm is given in Figure 6.

```
Statistical_Gate_Sizing(G)
1.   do {
2.       SSTA(G);
3.       For each gate x
4.           x.A'set = Initialize(x);
5.       gate_list = list of all gates x in G sorted by Sm_x
6.       Max_S = 0;
7.       while( gate_list is not empty) {
8.           x = Head(List);
9.           PropagateOneLevel(x);
10.          Update Sm_x ;
11.          if (x.curr_prop_level = # of levels in G) {
12.              gate_list = gate_list - {x};
13.              if (Max_S < S_x) {
14.                  Max_S = S_x;
15.                  best_gate = x;
16.              }
17.          }
18.          else
19.              Update position of x in gate_list;
20.          gate_list = gate_list - {x | Sm_x < Max_S}
21.      }
22.      best_gate.w = best_gate.w + Δw ;
23.  } while ( Max_S > 0) ;
```

Figure 6. Statistical Gate Sizing Algorithm

First, SSTA is performed to compute the arrival times at each node (step 2). To implement level by level propagation, we maintain propagated arrival times A' and $Sm_x$ for each candidate gate and use the notation '$x.A'set$' which represents the super-set of gates in the current perturbation front of gate $x$. It is a super-set as it also contains the fanout gates of the current perturbation front, which is required to advance the perturbation front one level forward. $Sm_x$ and A'set are initialized for every gate in the circuit by calling procedure **Initialize** (step 3 and 4). In step 5, a sorted list of all gates in G is created by arranging gates in descending order of $Sm_x$. It repre-sents the list of all unpruned candidates which may or may not result in maximum $S_x$, i.e. the set of all gates having $Sm_x > Max\_S$, where $Max\_S$ is the maximum $S_x$ amongst all candidate nodes whose perturbation front have reached the sink node. $Max\_S$ is initialized to be '0' (step 6) before beginning the search for the most sensitive gate.

In each iteration, the head of the sorted *gate_list* is selected for propagation. The procedure **PropagateOneLevel** propagates the arrival times by one level and updates the A'set of gate *x* accordingly, as explained later. During propagation, new nodes are added to A'set and nodes which do not belong to the perturbation front are deleted. $Sm_x$ is re-computed in step 10. If perturbation front of gate *x* reaches the sink node, gate *x* is removed from the candidate *gate_list* (step 11 and 12) and $Max\_S$ is updated (step 13-17). On the other hand, if the perturbation front has not yet reached the sink node, the position of gate *x* in the sorted *gate_list* is updated with respect to its new $Sm_x$ (step 19). In Step 20, gates in the list for which $Sm_x < Max\_S$ are removed from the list. When the candidate *gate_list* becomes empty the propagation loop terminates and the gate with maximum sensitivity is sized up by $\Delta w$. The algorithm can be easily modified to size multiple gates in the same iteration.

The pseudocode for procedure **Initialize** is given in Figure 7.

**Initialize(** gate x**)**
1. Change delays of x & fanin(x) for $\Delta w$ increase in x.w;
2. x.A'set = x $\cup$ fanin( x );
3. x.curr_prop_level = $\min_{i \in x.A'set}$ x.Aset[i].level;
4. while (x.curr_prop_level <= x.level)
5.    **PropagateOneLevel(x);**
6. Compute $Sm_x$ ;
7. Restore Change in delay of x & fanin(x);

**Figure 7. Initialize creates a perturbation front for gate x**

Given a gate *x*, the algorithm initializes the perturbation front A'set and computes the initial value $Sm_x$ for gate *x*. The width *w* of gate *x* is temporarily sized up by $\Delta w$ and the CDF for all pin-to-pin delays of gate *x* are updated. In addition, the delay CDFs of fanin gates that drive the inputs of gate *x* are also updated due to the increased loading by gate *x* (step 1). Consequently, the arrival time CDFs at gate *x* and its fanins will be perturbed and hence, these nodes are added to the initial perturbation front (step 2). In general, it is possible that the fanin of *x* is also a fanin of other fanins of *x*. Therefore, perturbed arrival times are computed starting from the minimum level gate (step 3). The procedure **PropagateOneLevel** is called iteratively until the propagation level reaches the level of gate *x* (step 4 and 5). The perturbation front created after step 5 is shown in Figure 8. Thereafter, $Sm_x$ is computed in step 6. Finally, the original pin-to-pin delays of gate *x* and its fanin gates are restored (step 7).

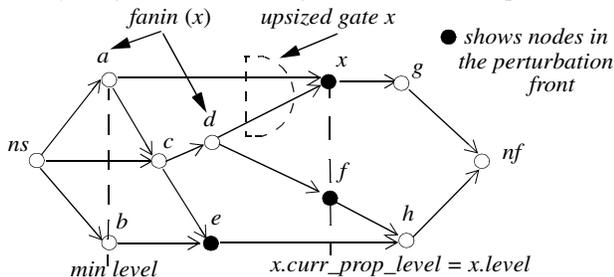

**Figure 8.** *Pertubation front* **after the Initialize routine for gate x**

The pseudocode for procedure **PropagateOneLevel** is given in Figure 9. In step 1-4, a list of gates at the propagation level is created from the A'set of x. For every node in the list, the perturbed arrival time distribution AT_PDF, is computed by performing convolution and max operations (step 6). Then $\Delta_i$ is computed by comparing the propagated perturbed arrival time and the original arrival time (i.e. the arrival time computed in step 2). For each node in prop_list, fanout nodes are added to the A'set in steps 8-12. Note that a node can be removed from the A'set once the perturbed arrival times at all its fanouts are computed. This is performed using a count variable *fo_count* which is initialized to the number of fanouts of the gate (step 11). When a perturbed arrival time is propagated from a node, *fo_count* of all fanin gates is decremented. A node is removed from A'set when its *fo_count* becomes zero.

**PropagateOneLevel(** gate x **)**
1. prop_list = $\phi$ ;
2. for_each_gate i $\in$ x.A'set
3.   if (i.level = x.curr_prop_level)
4.     prop_list = prop_list $\cup$ i;
5. for_each_gate i $\in$ prop_list {
6.   Compute x.A'set[i].AT_PDF using prop. & max;
7.   Compute $\Delta_i$ ;
8.   for_each_gate j $\in$ fanout( i )
9.     if (j $\notin$ x.A'set) {
10.      x.A'set = x.A'set $\cup$ j;
11.      x.A'set[j].fo_count = j.fo_count;
12.     }
13.   for_each_gate k $\in$ fanin( i )
14.     if (k $\in$ x.A'set) {
15.      x.A'set[k].fo_count = x.A'set[k].fo_count - 1;
16.      if ( x.A'set[k].fo_count = 0)
17.       x.A'set = x.A'set - { k };
18.     }
19. }
20. x.curr_prop_level = x.curr_prop_level + 1;

**Figure 9. Propagate One Level**

## 4 Results

The proposed statistical optimization method was implemented and tested on a synthesized version of ISCAS'85 [10] benchmark circuits using a 180nm commercial cell library. We compare our proposed approach with deterministic and brute-force statistical optimization methods. Intra-die process variation was modeled using a truncated Gaussian gate delay distribution. The standard deviation was 10% of the nominal delay and the distribution was truncated at the 3 sigma point. However, any delay distribution could be used in our framework. As shown later, the arrival time bounds as suggested in [3] were compared with Monte Carlo simulation, showing an acceptable difference, especially for the 99-percentile point (< 1%).

The deterministic optimization that we use for comparison is similar to a coordinate descent algorithm. Sensitivities are computed for all the gates on the critical path and the gate with the highest sensitivity is sized up. These sensitivities are computed as the change in the *circuit delay* due to a change in the gate size. The brute force statistical optimization, on the other hand, computes statistical sensitivities exactly by performing an SSTA run for every candidate gate.

Table 1 shows a comparison between the proposed statistical optimization and deterministic optimization. Results have not been shown for the brute force approach as they match exactly with the proposed optimization algorithm. The optimization results for the 99-percentile circuit delay point, after performing over 1000 sizing iterations, is shown in *column* 3-6. Note that while the deterministic optimization does not use SSTA in the optimization process, the reported 99-percentile delay point was obtained by running SSTA on the circuit solution after every sizing iteration. In *column* 3, we report the % increase in the total gate size of the circuit due to optimization. *Column* 4 and 5 show the 99-percentile delay obtained from deterministic and statistical optimization, respectively. The improvement obtained from statistical optimization over deterministic optimization is shown in *column* 6. The average improvement is 7.8% over all benchmarks with a maximum improvement of 10.5%.

**Table 1. Results for the 99-percentile delay point**

| Circuit | | Results for the 99-percentile delay pt. (ns) | | | |
|---|---|---|---|---|---|
| name | node/edge | % inc. | deterministic | statistical | %impr. |
| c432 | 214/379 | 97 | 3.49 | 3.14 | 10.03 |
| c499 | 561/978 | 25.6 | 3.98 | 3.56 | 10.55 |
| c880 | 425/804 | 93 | 4.09 | 3.74 | 8.55 |
| c1355 | 570/1071 | 23.7 | 4.80 | 4.30 | 10.41 |
| c1908 | 466/858 | 20.9 | 6.48 | 6.12 | 5.50 |
| c2670 | 1059/1731 | 21.4 | 3.65 | 3.40 | 6.85 |
| c3540 | 991/1972 | 11.5 | 5.98 | 5.70 | 5.0 |
| c5315 | 1806/3311 | 6.7 | 5.90 | 5.40 | 8.47 |
| c6288 | 2503/4999 | 28.1 | 16.00 | 15.05 | 5.93 |
| c7552 | 2202/3945 | 13.1 | 8.10 | 7.60 | 6.17 |

Table 2 shows a comparison of runtimes between brute force statistical optimization and our accelerated approach. Our optimization results are identical with those of the brute force approach, but provide a runtime improvement of up to 56$x$ for large circuits. The results demonstrate that as many as 55 out of 56 candidate nodes are pruned, demonstrating the effectiveness of the proposed bounds and pruning algorithm. We also found that the runtime per iteration varies significantly over the optimization iterations. In certain cases, there is a large range of gate sensitivities, and a highly sensitive gate quickly prunes out many inferior gates, whereas in other cases, many gates in the circuit have similar sensitivities, making pruning more difficult. Note that in the latter case, exact identification of the most sensitive gate will not be as important for the optimization result, opening the way for future research on fast heuristics for finding the most sensitive gate. In *column* 2 and 3, we report the average runtime per iteration (computed over a 1000 iterations) using the brute force and our accelerated approach, respectively. *column* 4 shows the runtime improvement factor. The improvement is higher for larger circuits, since in these circuits the cost of maintaining additional data structures is amortized over the savings of statistical computations, such as convolution and max. Finally, *columns* 5 and 6, show the range of runtimes per iteration for our approach and corresponding range of improvement factors, respectively.

Figure 10 shows the area-delay curve using our approach and deterministic optimization for c3540. The 99-percentile points of the circuit delay CDF are plotted on the x-axis and the corresponding total gate size value on the y-axis, for every sizing iteration. We have also plotted the 99-percentile points of the circuit delay using Monte Carlo simulations. As shown in the figure, there is a very small difference between the bounds and Monte Carlo results. Thus, using bounds as an optimization objective results in a comparable improvement in the exact circuit delay.

## 5 Conclusions

In this paper, we have demonstrated the need for a fast statistical optimization algorithm. We have shown through our experiments that there is a clear advantage in using statistical optimization com-pared to a deterministic one. We proposed a fast statistical optimization which is exact in comparison with a brute force statistical optimization but provides significant speedup. Our approach is based on proposed theory of perturbation bounds which allow efficient identification of the highest sensitive gate for sizing. The perturbation bound is used in pruning out less sensitive gates without explicitly propagating their effect, and is a key contribution of this paper. Finally, we demonstrated the accuracy and efficiency of our approach over a large number of test cases. Our results show a maximum runtime improvement by a factor of 56. Future work includes development of heuristics for fast and approximate identification of the statistically most sensitive gate in the circuit.

**Table 2. Results for the runtime improvement**

| Circuit name | Average time per iteration (sec) | | | Range of time per iteration(s) | Range of impr. factor |
|---|---|---|---|---|---|
| | brute force | our algo. | imp. factor | | |
| c432 | 5 | 1.35 | 3.7 | 0.72-1.81 | 3-7 |
| c499 | 90 | 22.4 | 4.01 | 5-30 | 3-18 |
| c880 | 15 | 4.0 | 3.75 | 1.5-5 | 3-10 |
| c1355 | 95 | 23 | 4.13 | 9-31 | 3-11 |
| c1908 | 102 | 25 | 4.08 | 10-36 | 3-10 |
| c2670 | 43 | 5.0 | 8.6 | 1.6-7.0 | 6-27 |
| c3540 | 194 | 28 | 6.9 | 6-35 | 6-32 |
| c5315 | 403 | 40 | 10.07 | 16-55 | 7-25 |
| c6288 | 3600 | 248 | 14.5 | 64-310 | 12-56 |
| c7552 | 1190 | 114 | 10.4 | 34-150 | 8-35 |

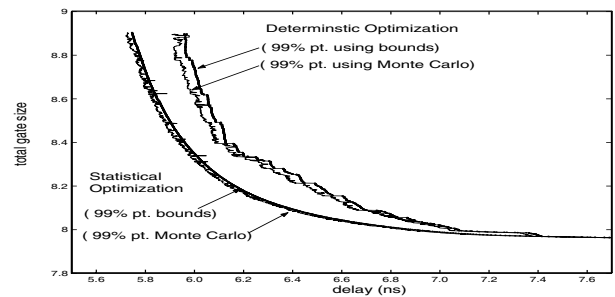

**Figure 10. Area- delay curve for c3540**

## Acknowledgements

This research was supported by SRC contract 2001-HJ-959 and NSF grant CCR-0205227.